\documentclass[referee]{mn2e}
\usepackage{graphicx}%
\usepackage{longtable}%
\usepackage{hhline}%

\title[Intense look at Virgo Southern Extension]
{Intense look at Virgo Southern Extension}

\author[I.\,D.\,Karachentsev, O.\,G.\,Nasonova]
{I.\,D.\,Karachentsev$^{1}$\thanks{E-mail:ikar@sao.ru},
O.\,G.\,Nasonova$^{1}$\thanks{E-mail:phiruzi@gmail.com}\\
$^{1}$Special Astrophysical Observatory of the Russian Academy
 of Sciences, Nizhnij Arkhyz, KChR, 369167, Russia}
\begin{document}


\pagerange{\pageref{firstpage}--\pageref{lastpage}} \pubyear{2012}

\maketitle

\label{firstpage}

\begin{abstract}
We collected data on radial velocities and distances of galaxies to elucidate
structure and kinematics of the filament attached to the Virgo cluster from
south. In the region $RA = [12.5 - 13.5]^h$, $Dec= [-20 - 0]^\circ$ there are
171 galaxies with radial velocities $V_{LG} < 2000$~km~s$^{-1}$, and 98 of them
have distance estimates. This galaxy cloud, called as ``Virgo Southern Extension'',
is situated just on the edge of the Virgo ``zero-velocity surface''. The mean
distance to Virgo SEx, $17\pm2$~Mpc, and the average radial velocity,
$1172\pm23$~km~s$^{-1}$, are very close to the Virgo cluster ones. In
Supergalactic coordinates the Virgo SEx dimensions are $15 \times 7 \times
2$~Mpc, where the major axis is directed along the line of sight, the
second-major axis looks towards the Virgo core and the minor one is
perpendicular to the Supergalactic plane. This flattened cloud consists of a
dozen virialized groups with the total $K$-band luminosity of $1.7\cdot10^{12}
L_\odot$ and the total virial mass of $6.3\cdot10^{13} M_\odot$, having a
typical dark matter-to- stellar matter ratio of 37. The Hubble diagram for
Virgo SEx galaxies exhibits a tendency of $Z$-shape wave with a velocity amplitude of
$\sim250$~km~s$^{-1}$ that may be caused by a mass overdensity of
$\sim6\cdot10^{13} M_\odot$, and in order of magnitude agrees with 
the sum of virial masses of the groups.
\end{abstract}


\section{Introduction}

The virial mass estimates for groups and clusters of galaxies in the Local
Universe within $\sim50$ Mpc result in the average matter density $\Omega_m\simeq
0.06-0.10$ (Vennik 1984, Tully 1987, Magtesian 1988, Makarov \& Karachentsev
2011), which is 3--4 times less than the global value $\Omega_m\simeq 0.25-0.30$
(Spergel et al. 2007). Several possible explanations of this discrepancy were regarded in
the overview of (Karachentsev 2012). The ``missing dark matter'' paradox can be
sourced by the presence of extended dark ``suburbs'' distributed mainly outside
the virial radii of groups and clusters of galaxies (Tavio et al. 2008, Masaki
et al. 2011, Chernin et al. 2012). Some authors suggest that the hidden dark
matter is concentrated in dark filaments and bridges between clusters
(Dinshaw et al. 1997, Impey et al. 1999, Rosenberg et al. 2003). The first
observational evidence of their existence has appeared recently (Dietrich et al.
2012).

The Virgo cluster of galaxies situated at a distance of 16.5 Mpc (Mei et al. 2007) is the
nearest cluster with an angular diameter of the virial core $\sim12^{\circ}$. It
provides a unique possibility to search for dark matter structures. Karachentsev
et al. (2011, 2012) discussed the kinematics of two regions: Coma~I and
Ursa~Majoris, extending north from the Virgo cluster along the Supergalactic
equator. The UMa region has turned to be an association of common groups with
the normal virial mass-to-luminosity ratio. However, the observed ``$Z$-wave''
disturbance of the Hubble flow in the Coma~I region possibly points out the
existence of a dark attractor with the mass of $\sim1\cdot10^{14}M_{\odot}$
situated 15 Mpc from us.

It is well known that bright galaxies are concentrated along the Supergalactic
equator not only northward of the Virgo cluster but also southward (de
Vaucouleurs \& de Vaucouleurs 1973). Tully (1982) has called this structure
``Virgo Southern Extension''. According to Tully (1982), Virgo SEx is
characterized by dimensions of $\Delta SGX\simeq\Delta SGY\simeq10$~Mpc and
$\Delta SGZ\simeq4$~Mpc, bordering directly on the virial core of the Virgo
cluster. Virgo SEx occupies approximately a region of sky given by $RA =
[12.0^h-13.5^h]$, $Dec = [+5^{\circ}, -20^{\circ}]$, while the radial velocities
of galaxies populating it belong to the interval [$500<V_{LG}
<1700]$~km~s$^{-1}$. The Virgo SEx location proposed by Tully seems to be rather
conventional. Other authors (Binggeli et al. 1993, Yoon et al. 2012) limited the
Virgo Southern Extension by the interval $Dec\simeq [0, +5^{\circ}]$.
Considering the distance estimates for 22 galaxies in the Virgo SEx area, Tully
\& Shaya (1984) inferred that this filament guides galaxy flow into the virial
zone of the Virgo cluster. Yoon et al. (2012) examined the absorption
$\L_{\alpha}$-lines in spectra of 7 quasars situated behind the Virgo SEx
and suggested the existence of a large-scale flow of warm intergalactic gas
towards the cluster centre.

It is worth emphasizing that structure and kinematics of the Virgo Southern
Extension lack for researchers' application until recently. The accumulated
observational data allow now to investigate this complex of galaxies in more detail.

\section{Structure of the Virgo Southern Extension region}

Optical and HI surveys carried out recently have improved significantly the
observational basis for analysis of Virgo SEx structure. We restricted our
consideration to the sky region with equatorial coordinates $RA =
[12^h30^m-13^h30^m]$, $Dec = [-20^{\circ}, 0^{\circ}]$ and to the radial
velocities relative to the Local Group centroid $V_{LG}<2000$~km~s$^{-1}$.
According to the catalogue of nearby galaxy groups (Makarov \& Karachentsev
2011) there is about a dozen groups in this area with corresponding velocities.
A list of 171 galaxies in the Virgo SEx zone with $V_{LG}<2000$~km~s$^{-1}$ is
represented in Table~1. It includes several galaxies outside the mentioned
area which belong to the groups with centres lying inside the region. The
columns of the table contain: (1) galaxy name in known catalogues; (2) its
equatorial coordinates at the epoch J2000.0; (3) radial velocity in the LG
reference frame; the main sources of velocity data are LEDA
(http://leda.univ-lyon1.fr) and NED (http://ned.ipac.caltech.edu) databases; (4)
morphological type on the de Vaucouleurs scale; most types were determined by us
independently of other sources; (5) apparent magnitude of a galaxy in the
$K_s$-band from 2MASS survey (Jarrett et al. 2000); the 2MASS lacks
$K_s$-data for many blue and diffuse galaxies, so in these cases we make
estimates for $K_s$-magnitude from $B$-magnitude and mean colour index $<B-K>$
for galaxies of each morphological type according to procedure described by
Jarrett et al. (2000); (6) name of the dominating galaxy of a group according to
Makarov \& Karachentsev (2011); (7,8) distance modulus and corresponding
distance (in Mpc); distance estimates are taken from NED favouring most recent
data (Tully et al. 2009; Springob et al. 2009); in more than half the instances
we estimated (M--m)- moduli ourself via Tully-Fisher relation using data on
apparent magnitudes $B_T$ and HI line widths $W_{50}$; (9) method applied for
estimating distance: ``tf''~-- from Tully-Fisher relation, ``sbf''~-- from
surface brightness fluctuations (Tonry et al. 2001), ``SN''~-- from Supernovae
luminosity, ``mem''~-- from galaxy membership in a group with measured distance.

Galaxy distribution in the Virgo SEx region is represented in Fig.~1 with
circles. Large circles mark most bright galaxies with apparent magnitudes
$K_s<9.0^m$. In the upper panel of the figure galaxies of early, intermediate and
late morphological types are shown with different colours. The solid line
corresponds to the Supergalactic equator. In the lower panel of the figure
colours distinguish galaxies in different velocity ranges, while group members
are linked with dominating galaxies by straight lines.

\begin{figure}
\includegraphics[height=0.6\textwidth,keepaspectratio,angle=270]{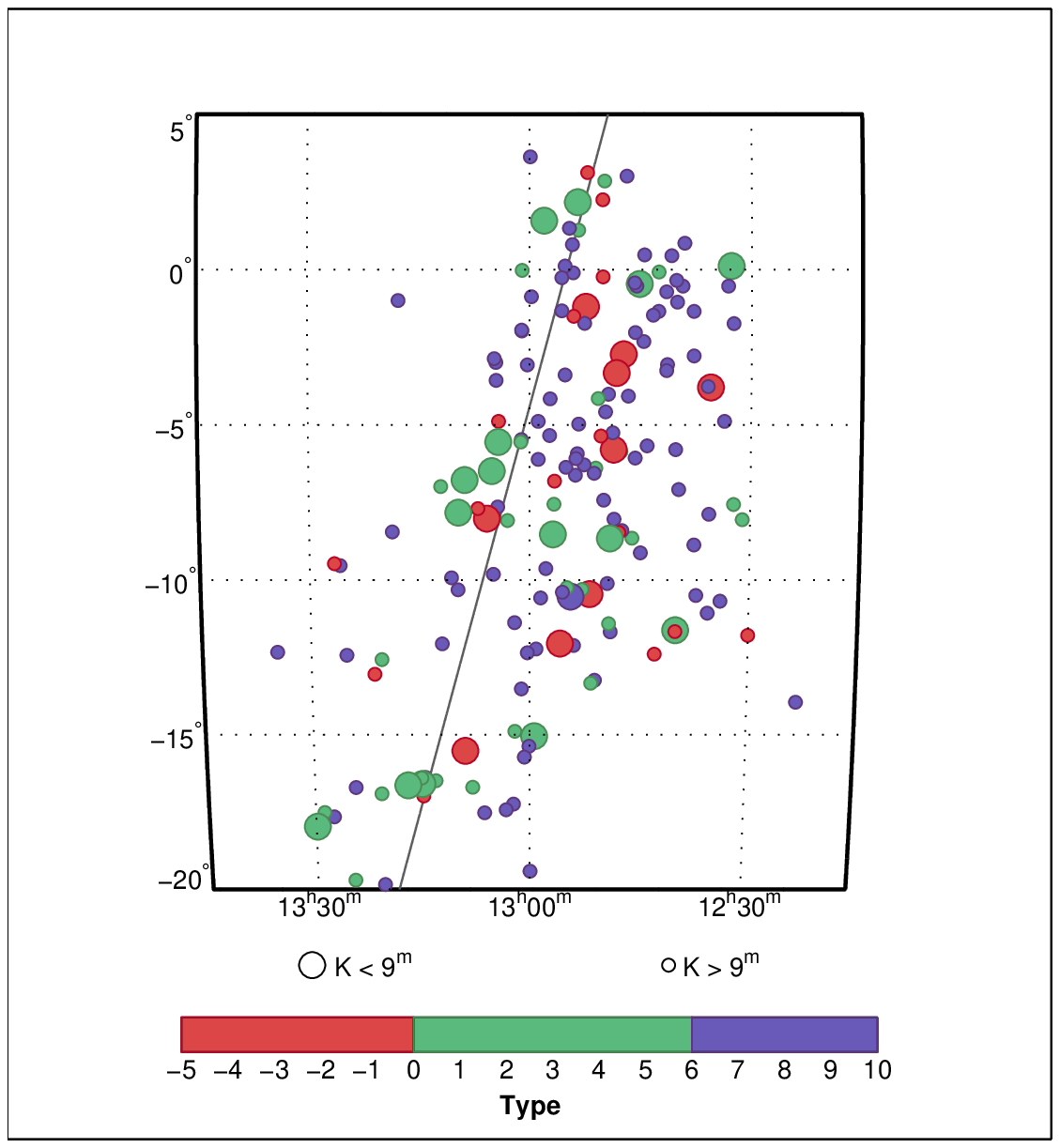}\\
\includegraphics[height=0.6\textwidth,keepaspectratio,angle=270]{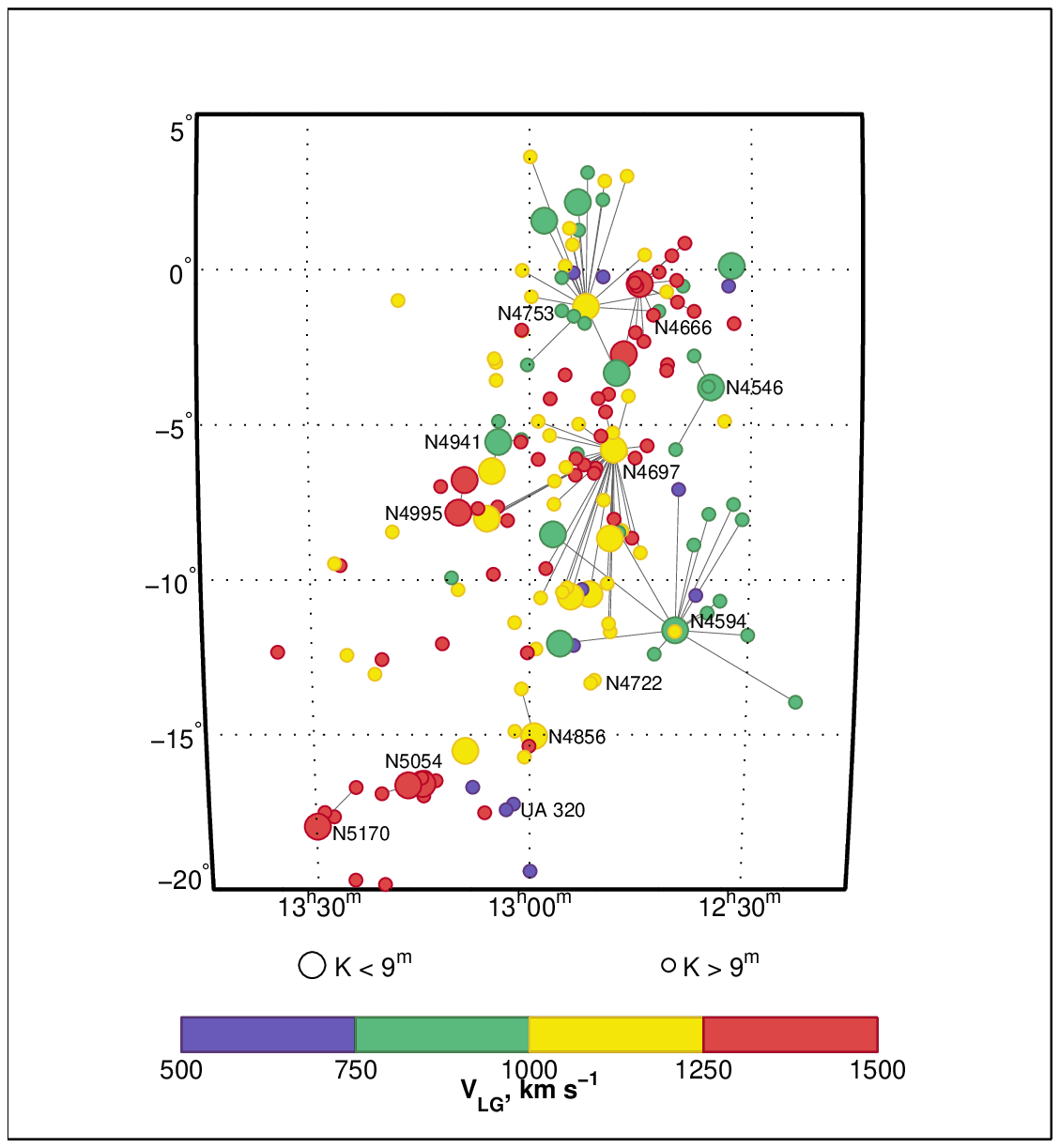}
\caption{The Virgo Southern Extension complex of galaxies in equatorial
coordinates. {\itshape{}Top:} early type, intermediate type and late type
galaxies indicated by different colours. Bright ($K < 9^m$) galaxies are shown
by larger circles. {\itshape{}Bottom:} the same field with indication of radial
velocities of the galaxies and their membership in different MK-groups.}
\end{figure}

The total distribution of galaxy number over radial velocities for Virgo SEx
complex is shown in Fig.~2; the galaxies with prevailing bulge component ($T<3$)
are greyed.

\begin{figure}
\includegraphics[width=0.8\textwidth,keepaspectratio]{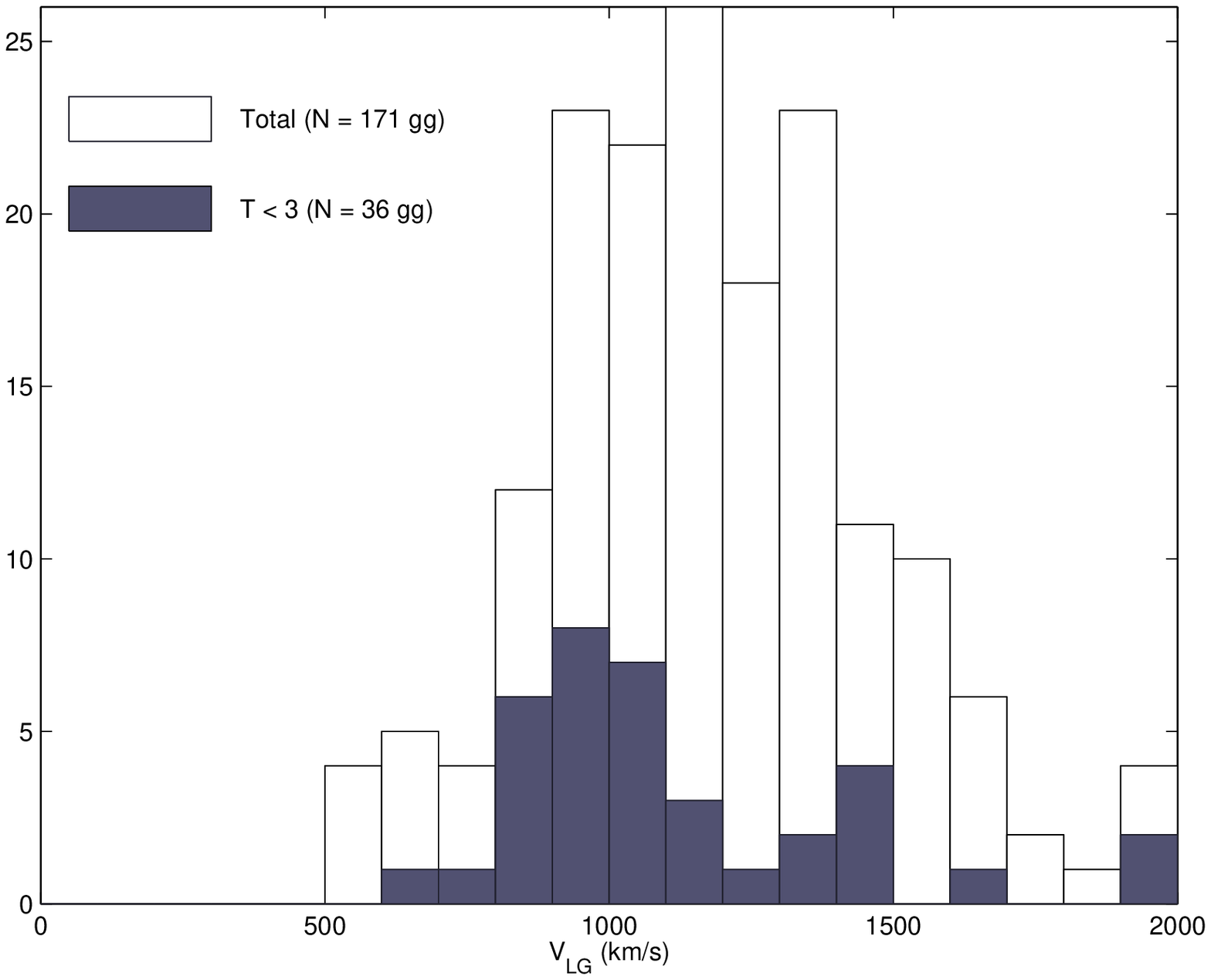}
\caption{Distribution of galaxies in the Virgo SEx region on their radial
velocities in the Local Group frame. Galaxies with prominent bulges are shaded.}
\end{figure}

According to this data the distribution of galaxies over the sky seems to be
clumpy and extended along the Supergalactic equator. The mean radial velocity
and the mean morphological type of galaxies are actually independent on
galaxies' distance from the Virgo centre. The total distribution of Virgo SEx
galaxies over radial velocities follows in good agreement the Gaussian shape with
the mean $<V_{LG}>=+1172$~km~s$^{-1}$ and the standard deviation
$\sigma_v=285$~km~s$^{-1}$. For galaxies with prominent bulges these values are
1106~km~s$^{-1}$ and 282 km s$^{-1}$, respectively. The observed distribution
$N(V_{LG}$) differs significantly from a strictly increasing one predicted by a
uniform distribution of galaxies in the space. In other words, the filamentary
structure of Virgo SEx can be considered as a kinematically segregated
formation.

Some specific features of the sky region under investigation should be noted. It
follows from the data by Tully \& Shaya (1984) and Karachentsev \& Nasonova
(2010) that the radius of the zero velocity surface around the Virgo cluster is
$R_0\simeq6.5$~Mpc or $\sim23^{\circ}$. Therefore, the southern boundary of the
accretion sphere passes just through the centre of the Virgo SEx complex along
the line of $Dec\simeq -10^{\circ}$. Alternatively, the Virgo SEx area is
situated not far from the Great Attractor direction, so the kinematics of the
considered sample of galaxies can be influenced by this aggregate. Unfortunately
the zone which includes the Virgo SEx region is not represented photometrically
in the CGCG (Zwicky et al. 1961--1968) and ESO (Lauberts \& Valentijn 1985)
catalogues. Even some bright galaxies populating this area have therefore a
certain disagreement in their apparent magnitude estimates to an extent of
2$^m$--3$^m$. However, a part of considered region is already covered with SDSS
data (Abazaijan et al. 2009) and, moreover, this structure is located completely
in the HIPASS survey zone (Zwaan et al. 2003, Meyer et al. 2004).

\section{Hubble flow in the Virgo Southern Extension complex}

The velocity-distance relation for 98 galaxies in the Virgo SEx zone is
represented in the upper panel of Fig.~3. The straight line corresponds to the
unperturbed Hubble flow with the parameter $H_0=73$~km~s$^{-1}$Mpc$^{-1}$.
Isolated galaxies are shown as crosses while the members of different groups
marked with circles are linked with dominating galaxies by straight lines. The
triangles indicate 26 members of the NGC\,4697 group which seems to be an
artificial group (see Section~4 for details). The thick broken line
demonstrates the behaviour of the running median with a window 1~Mpc.

\begin{figure}
\includegraphics[height=0.8\textwidth,keepaspectratio,angle=270]{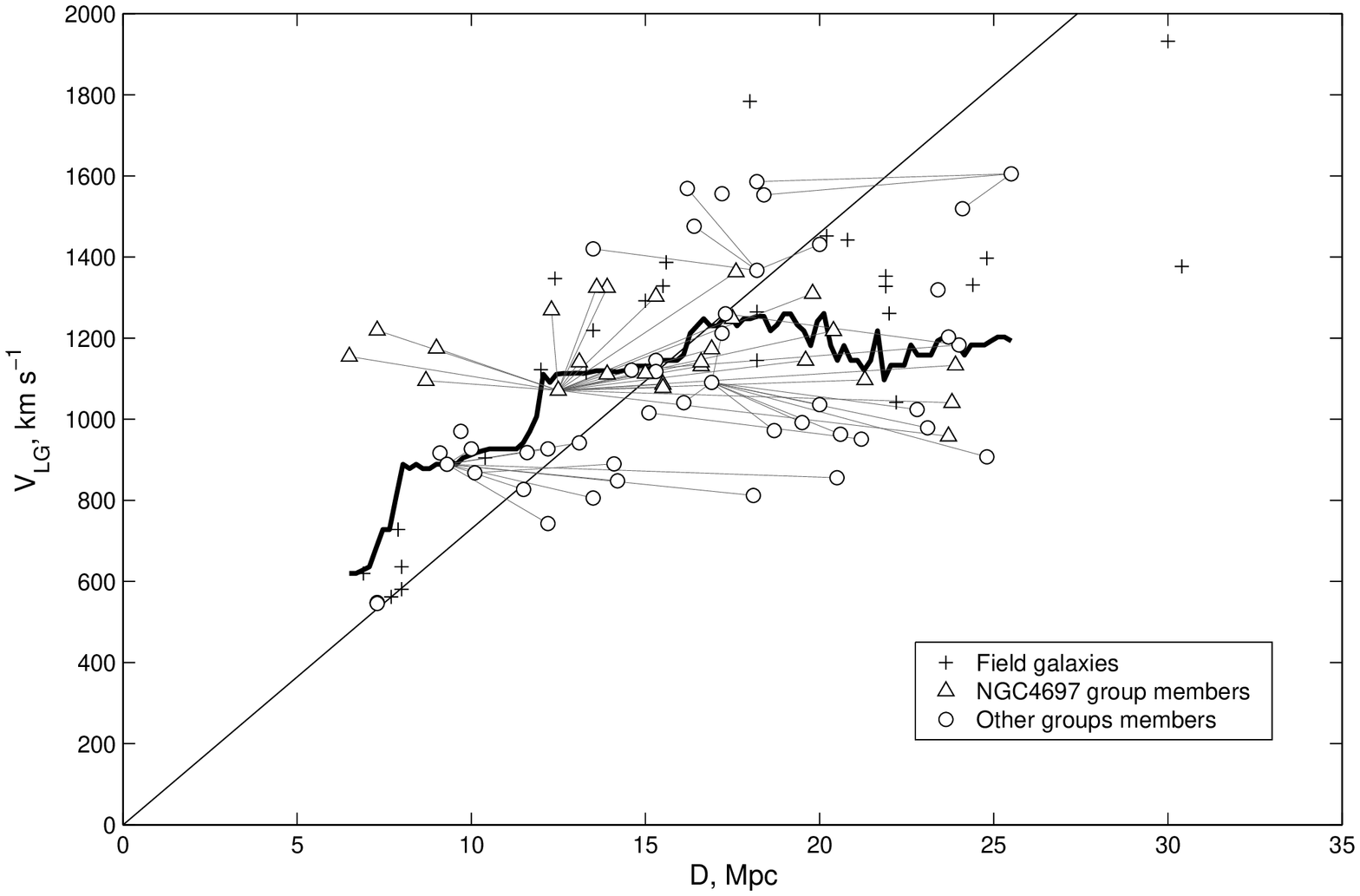}
\includegraphics[height=0.8\textwidth,keepaspectratio,angle=270]{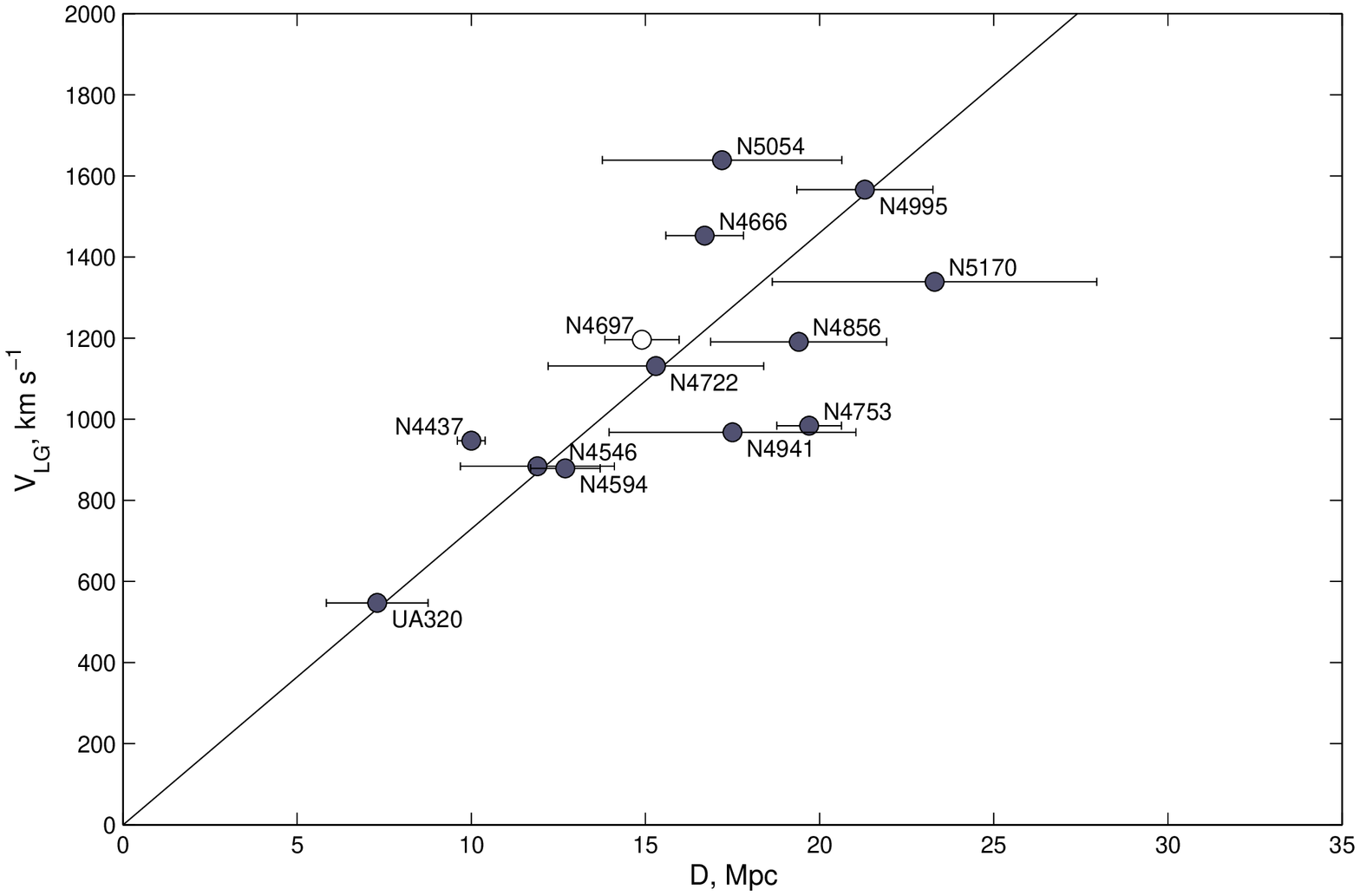}
\caption{The velocity-distance relation for galaxies in the Virgo Southern
Extension area. The undisturbed linear Hubble flow with $H_0 =
73$~km~s$^{-1}$Mpc$^{-1}$ is shown by the straight line. {\itshape{}Top:}
individual galaxies; the solid polygon curve traces the running median on
observational data with a window of 1~Mpc. {\itshape{}Bottom:} centres of the
groups.}
\end{figure}

Note some features of this diagram.

As a first approximation, both field galaxies and groups members follow the
Hubble relation with the slope mentioned above (the $H_0$ parameter). About
90\,\% of all galaxies have distance estimates in the range from 8 to 25~Mpc.
Taking into account the typical distance error of $\sim20$\%, the length of the
Virgo SEx structure along the line of sight reaches about 15~Mpc which is twice
more then its projected dimensions on the sky (7~Mpc). The mean distance to
Virgo SEx galaxies, 17~Mpc, is roughly the same that the distance to Virgo
cluster centre, i.e. the Virgo SEx filament is attached to the cluster at nearly right
angle with respect to the line of sight. The behaviour of the running median indicates
the presence of $Z$-shaped wave above the unperturbed Hubble flow
with an amplitude of $\sim(200-300)$~km~s$^{-1}$ that may be caused by
a mass overdensity. As it can be seen from data
by Tully \& Shaya (1984) (see their Fig.4), the amplitude of $Z$-wave caused by
infall of galaxies towards the cluster mass $\sim7\cdot10^{14}M_{\odot}$ should be only
(30--90)~km~s$^{-1}$ at angular distances
21$^{\circ}$--26$^{\circ}$ from the Virgo centre. Hence, the observed amplitude
$\sim250$ km s$^{-1}$ may be due to the collapse of the Virgo SEx complex
itself. In the simplest model of the Hubble flow around point mass overdensity
such amplitude corresponds to a mass of $\sim6\cdot10^{13}M_{\odot}$.

The lower panel of Fig.~3 represents relation between mean distances and
mean velocities for the centres of 13 groups. Each group is marked with a circle
and labelled with the name of its brightest galaxy. Distance error bars are
marked by horizontal lines. As one can see, most groups in the Virgo SEx region
follow the unperturbed Hubble flow with $H_0=73$~km~s$^{-1}$Mpc$^{-1}$ quite
well within the error of galaxy distances. However, the NGC\,4437 and NGC\,4666
galaxy groups belonging to the foreground of the complex show signs of
infall towards the complex centre. The NGC\,4753 group demonstrates a
significant effect of infall from the back side of the Virgo Southern Extension.

\section{Total mass of the Virgo Southern Extension complex}

The main characteristics of 13 groups in the considered region of sky
according to Makarov \& Karachentsev (2009, 2011) and Karachentsev \& Makarov
(2008) are presented in Table~2. The table columns contain: (1) name of
the dominating galaxy, (2) equatorial coordinates of the group centre, (3)
number of members with measured radial velocities, (4) mean radial velocity in
the LG reference frame (km~s$^{-1}$), (5) dispersion of radial velocities
(km~s$^{-1}$), (6) mean harmonic radius (kpc), (7) logarithmic total luminosity
in the $K_s$-band in solar units, (8) logarithm of virial (projected) mass (in $M_{\odot}$),
estimated as  $M_p = \frac{32}{\pi G} (n - 3/2)^{-1} \sum\limits_{i=1}^n V_i^2 R_i$,
where $V_i$ and $R_i$ are radial velocity and projected distance of the $i$-th
galaxy relative to the system centre (Heisler et al. 1985),
(9) virial mass to $K$-luminosity ratio in solar units, (10) morphological type
of the dominating galaxy, (11, 12) mean distance modulus and the group member modulus variance, (13)
distance to the group centre (Mpc) corresponding to the mean modulus, (14) number of
group members with measured distances. The initial mean parameters for these
groups were slightly changed due to revised distances and appearance of new
measurements of radial velocities.

It follows from the last column data of Table~1 that most galaxies in the
considered area have distances estimated from Tully-Fisher relation (1977) with
an accuracy of $\sim20$\% or $\sim0.4^m$. All the groups in Table~2, besides the
NGC\,4697 group, have ($m-M$) estimates dispersion of the same order: the rms
value $\sigma(m-M)$ for the sample of groups is just 0.40$^m$. This result
confirms the efficiency of algorithm used for selecting groups of galaxies
(Makarov \& Karachentsev 2011). The only exception is the 
NGC\,4697 group. It presents a puzzle being characterised by small
dispersion of radial velocities while the individual distance estimates of its
members are spread widely from 7 to 24 Mpc. The galaxy grouping criterion
(Makarov \& Karachentsev 2011) uses radial velocities of galaxies as an
estimator of their distances. In some regions of the sky where the velocity
field is distorted by a local matter overdensity the grouping
algorithm is constrained as its application can result in selecting artificial
groups containing field galaxies and members of their associations which are not
physically bound. Supposedly NGC\,4697 and other 35 galaxies with similar
velocities compose probably such a kinematically spurious ``pseudogroup''.

The total $K$-luminosity of galaxy groups forming the Virgo SEx complex reaches
$1.7\cdot10^{12}L_{\odot}$ while field galaxies contribute only 10\,\% of this
value. The sum of virial mass estimates for 13 groups from Table~2 amounts to
$6.3\cdot10^{13}M_{\odot}$ which is practically similar to the estimate of the
Virgo SEx total mass, $6\cdot10^{13}M_{\odot}$, obtained afore from the infall
amplitude. This agreement between both total mass estimates for the Virgo SEx
complex made from internal and external motions looks notable. The total
mass to $K$-luminosity ratio of the complex is $37\cdot M_{\odot}/L_{\odot}$.
The global matter density $\Omega_m=0.28$ in the Standard model with $H_0=73$~km~s$^{-1}$Mpc$^{-1}$
(Fukugita \& Peebles, 2004) and the mean K-band luminosity density $j_K = 4.3\cdot10^8 L_{\odot} Mpc^{-3}$
(Jones et al. 2006) correspond to the ratio $M/L_K=97\cdot M_{\odot}/L_{\odot}$. 
Therefore, the obtained ratio of total mass-to-total K-luminosity,
$37\cdot M_{\odot}/L_{\odot}$, indicates the presence of a moderate
amount of dark matter in the Virgo SEx complex
with a typical value of $\Omega_m\simeq0.11$.

According to data presented in Table~2, the NGC\,4594~= M\,104~= ``Sombrero'' group with
mass-to-luminosity ratio $M/L_K=38\cdot M_{\odot}/L_{\odot}$ is among 
the most ``dark'' substructures in the
Virgo SEx region. This group resides
in the outskirts of the Local Volume and is populated with some faint dwarf
galaxies of low surface brightness (Karachentsev et al. 2013),
whose radial velocities and distances are not measured yet. Note, that
almost all companions to Sombrero have distances, estimated via TF-relation, a bit higher
than that of the Sombrero itself, estimated from surface brightness fluctuations 
(Tonry et al. 2001). A reason of this difference is not clear to us. 
It may be caused by an accidental error in the sbf- distance to Sombrero.
Determination of accurate distance to it by the tip of red giant branch method with
ACS HST would be a cardinal way to solve the problem.

\section{Concluding remarks}

\begin{figure}
\includegraphics[height=0.6\textwidth,keepaspectratio]{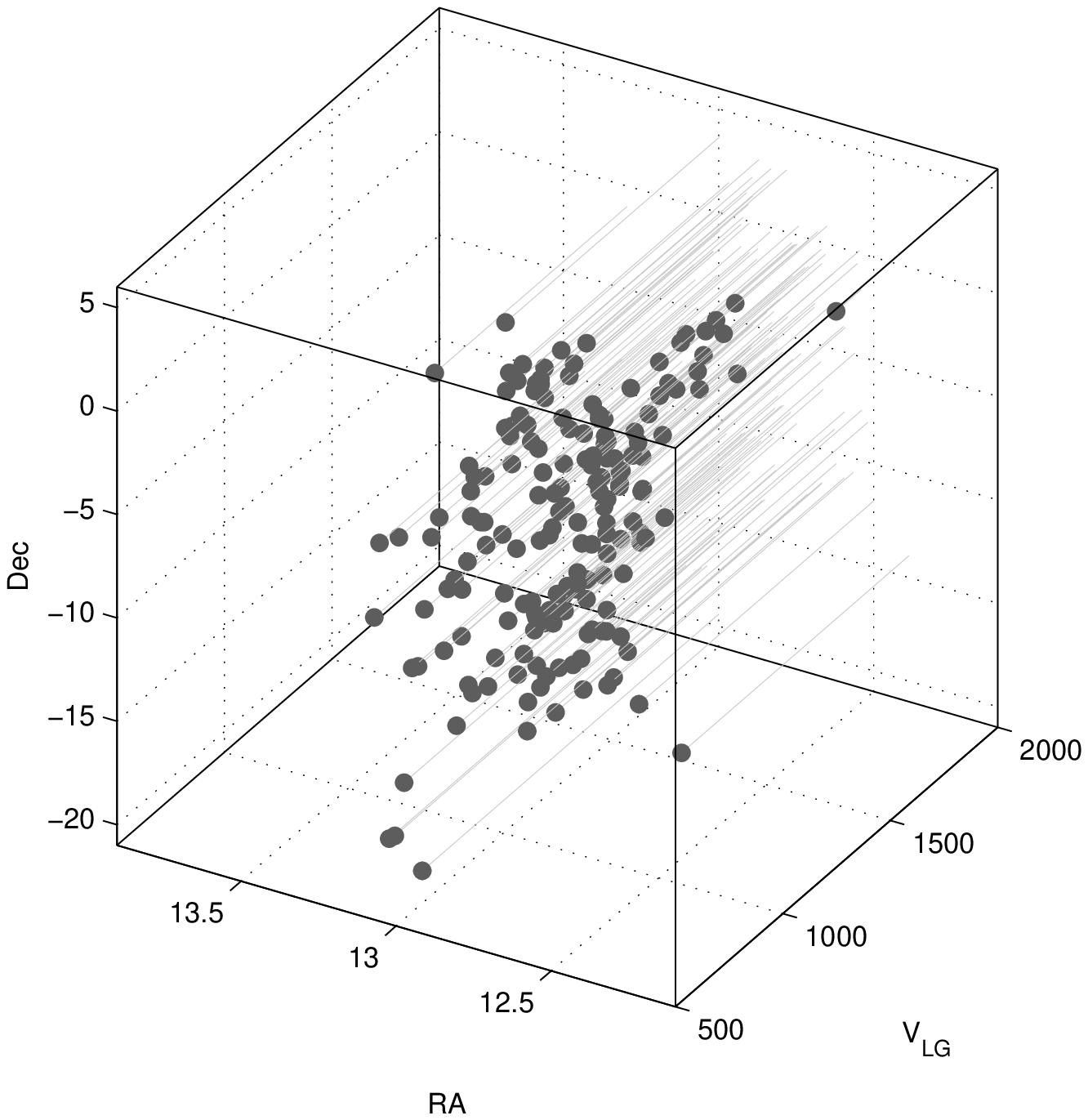}\\
\includegraphics[height=0.6\textwidth,keepaspectratio]{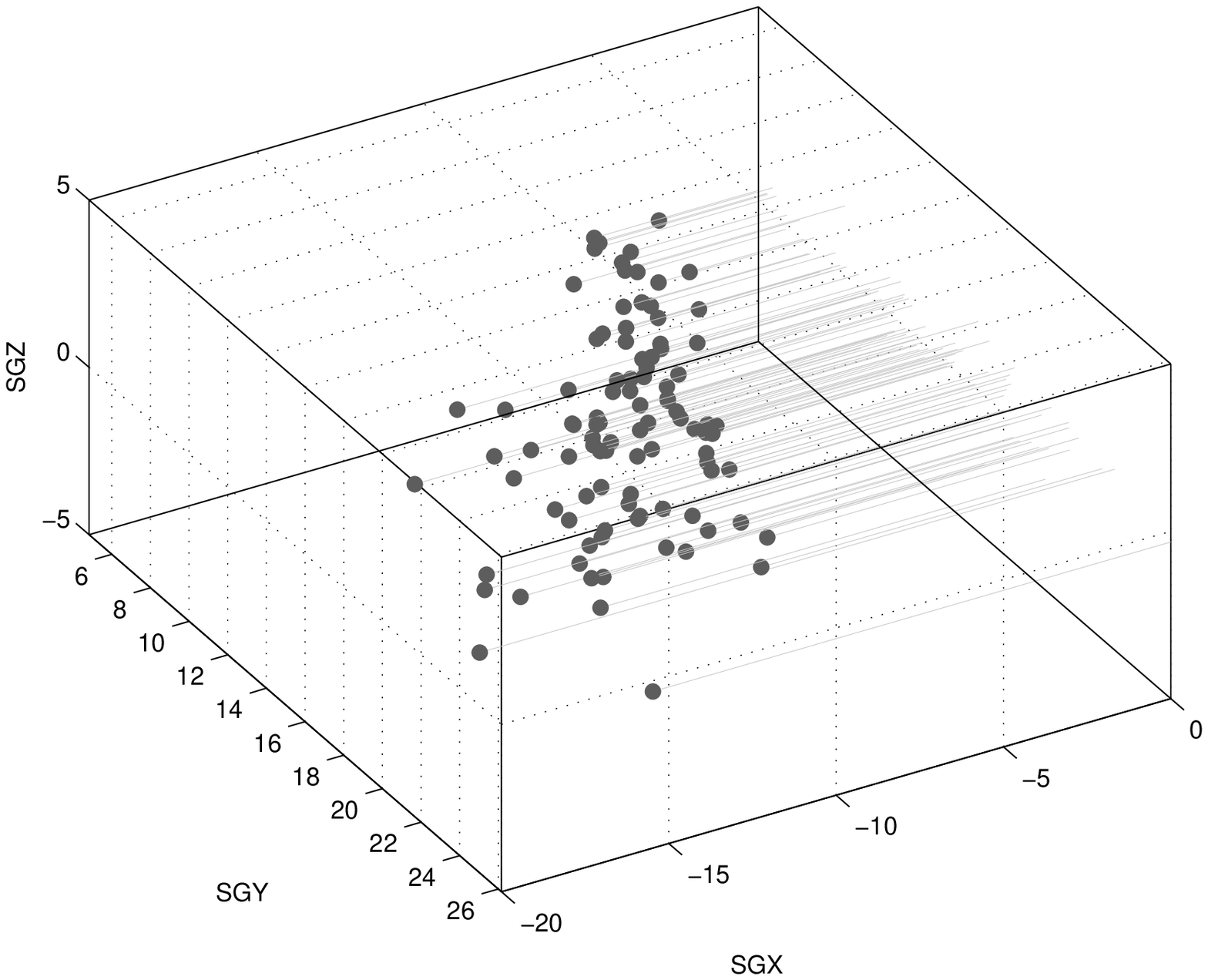}
\caption{Overview of the Virgo SEx structure. {\itshape{}Top:} galaxies with
measured radial velocities. {\itshape{}Bottom:} galaxies with individual
distance estimates in Supergalactic coordinates.}
\end{figure}

The Virgo Southern Extension, an extended complex of galaxies situated south of
the Virgo cluster, seems to be one of the closest examples of cosmic
filaments which appear commonly in N-body simulated large scale structure of the
Universe. In the considered region of the sky given by $RA =
[12^h30^m-13^h30^m]$, $Dec = [0, -20^{\circ}]$ there are 171 galaxies with
radial velocities $V_{LG}<2000$~km~s$^{-1}$, among them 98 galaxies have
individual distance estimates. In Supergalactic coordinates the Virgo SEx
structure is roughly characterized by dimensions of $\Delta SGX=15, \Delta
SGY=7$ and $\Delta SGZ=2$~Mpc where the major axis is directed along the line of
sight, the second-major axis looks towards the Virgo cluster and the minor axis
is perpendicular to the Supergalactic plane (see Fig.~4). Hence, the Virgo SEx structure
looks like a filament only in projection, having the true shape of a sheet.

Some authors claim that the Virgo SEx is somewhat like a cosmic feeding pipe
conducting galaxies and intergalactic gas into the virialized core of the Virgo
cluster (Tully \& Shaya 1984, Yoon et al. 2012). Nevertheless, the mean radial
velocity of Virgo SEx galaxies, $<V_{LG}>=(1172\pm23)$~km~s$^{-1}$, and their
mean distance, $<D>=(17\pm2)$~Mpc, are essentially the same as the corresponding
values for the Virgo cluster itself. Taking this into account, the radial flow
of matter towards the Virgo cluster could be hardly registered observationally.

About 80\,\% of galaxies populating the Virgo SEx region are members of the
MK-groups (Makarov \& Karachentsev 2011). The total $K$-luminosity of these
groups amounts to $1.7\cdot10^{12}L_{\odot}$ while their total virial mass
reaches $6.3\cdot10^{13}M_{\odot}$. As the mean $K$-band luminosity of stars
matches their mean mass, the ratio of dark matter to visible (stellar) matter for
the Virgo Southern Extension as a whole is $M_{DM}/M_*=37$, i.e. a bit less than
the corresponding value, $48\pm6$, for the Virgo cluster (McLaughlin,1999).

The velocity-distance diagram for 98 galaxies of the Virgo SEx area shows some
signs of $Z$-shaped wave probably caused by the deceleration of the Hubble flow due
to the Virgo SEx overdensity. The observed amplitude of this effect
$\sim250$~km~s$^{-1}$ corresponds to the overdensity mass of
$\sim6\cdot10^{13}M_{\odot}$ being in agreement with the sum of virial
masses of 13 MK-groups found in this region.

{\bf Acknowledgements}

\noindent{}We thank Dmitry Makarov for valuable comments.
This work was supported by the Russian Foundation for Basic Research (grants
10-02-00123, 11-02-00639, 11-02-90449, 12-02-91338-DFG), DFG grant G0563/23-1 and a grant of the
Ministry of Education and Science of the Russian Federation (No 14.740.11.0901).
O.\,G.\,Nasonova thanks the non-profit Dmitry Zimin's Dynasty Foundation for the
financial support.

{}

\begin{center}
\begin{longtable}{lcrrrcccl}
\caption{List of galaxies in the Virgo SEx region}\\
\hhline{=========}
 Galaxy       & RA (j2000) Dec.  &$V_{LG}$ & $T$ & $m_K$ &  group   & $m$-$M$  &   $D$  &note  \\
\hhline{=========}\endfirsthead
\hhline{=========}
 Galaxy       & RA (j2000) Dec.  &$V_{LG}$ & $T$ & $m_K$ &  group   & $m$-$M$  &   $D$  &note  \\
\hhline{=========}\endhead
\hhline{---------}\endfoot
\hhline{=========}\endlastfoot
UGCA 278          & 122310.3$-$135645 & 942 &  10 &13.6 & NGC4594  &30.58   &       13.1   & tf   \\
2MASX ...         & 123001.8$-$114731 & 909 &$-$0 &13.3 & NGC4594  &  $-$   &        $-$   &      \\
CGCG014-054       & 123103.8$+$014033 & 954 &   9 &13.2 & NGC4437  &29.91   & \enskip9.6   & tf   \\
NGC4487           & 123104.5$-$080314 & 848 &   6 & 9.3 & NGC4594  &30.77   &       14.2   & tf   \\
NGC4504           & 123217.5$-$073348 & 812 &   6 & 9.5 & NGC4594  &31.29   &       18.1   & tf   \\
UM 504            & 123223.6$-$014424 &1883 &   9 &13.8 &          &  $-$   &        $-$   &      \\
NGC4437\,=\,N4517 & 123245.6$+$000654 & 970 &   6 & 7.3 & NGC4437  &29.94   & \enskip9.7   & tf   \\
KDG155            & 123308.0$-$003159 & 581 &  10 &14.0 &          &  $-$   &        $-$   &      \\
UGCA 286          & 123337.0$-$045306 &1115 &   8 &10.9 &          &30.62   &       13.3   & tf   \\
UGCA 287          & 123355.2$-$104048 & 856 &   8 &11.3 & NGC4594  &31.56   &       20.5   & tf   \\
NGC4546           & 123529.5$-$034735 & 890 &$-$3 & 7.4 & NGC4546  &30.74   &       14.1   & sbf  \\
UGCA 289          & 123537.5$-$075240 & 806 &   8 &11.1 & NGC4594  &30.65   &       13.5   & tf   \\
PGC0970397        & 123539.4$-$110402 & 927 &  10 &13.2 & NGC4594  &30.01   &       10.0   & tf   \\
CGCG 014-074      & 123551.0$-$034558 & 829 &   9 &12.5 & NGC4546  &  $-$   &        $-$   &      \\
$[$KKS2000$]$ 29  & 123714.1$-$102951 & 562 &  10 &14.1 &          &29.45   & \enskip7.7   & tf   \\
$[$KKS2000$]$ 30  & 123735.9$-$085202 & 917 &  10 &14.8 & NGC4594  &29.79   & \enskip9.1   & tf   \\
LCRS B...         & 123746.5$-$024654 & 947 &   9 &13.7 & NGC4546  &  $-$   &        $-$   &      \\
CGCG 014-080      & 123748.3$-$012042 &1377 &   9 &13.5 & NGC4666  &  $-$   &        $-$   &      \\
SDSS ...          & 123902.5$+$005059 &1436 &   9 &12.2 & NGC4666  &  $-$   &        $-$   &      \\
NGC4592           & 123918.7$-$003155 & 918 &   8 &10.2 & NGC4437  &30.17   &       11.6   & tf   \\
HIPASSJ1239-07    & 123944.5$-$070514 & 743 &  10 &15.1 & NGC4594  &30.44   &       12.2   & tf   \\
NGC4594           & 123959.4$-$113723 & 889 &   1 & 4.9 & NGC4594  &29.85   & \enskip9.3   & sbf  \\
SDSS ...          & 124002.8$-$010300 &1423 &  10 &13.3 & NGC4666  &  $-$   &        $-$   &      \\
SUCD1             & 124003.1$-$114004 &1097 &$-$1 &14.7 & NGC4594  &  $-$   &        $-$   &      \\
SDSS ...          & 124008.9$-$002101 &1546 &  10 &13.1 & NGC4666  &  $-$   &        $-$   &      \\
NGC4597           & 124012.9$-$054757 & 868 &   8 &11.7 & NGC4546  &30.01   &       10.1   & tf   \\
SDSS ...          & 124048.5$+$002655 &1616 &   9 &15.3 & NGC4666  &  $-$   &        $-$   &      \\
CGCG 014-104      & 124122.9$-$030329 &1292 &   8 &11.4 &          &30.89   &       15.0   & tf   \\
FGC 1485          & 124128.9$-$031513 &1642 &   8 &14.6 &          &  $-$   &        $-$   &      \\
SDSS ...          & 124129.2$-$004311 &1027 &   9 &15.9 & NGC4753  &  $-$   &        $-$   &      \\
NGC4629           & 124232.7$-$012102 & 963 &   9 &11.8 & NGC4753  &31.56   &       20.6   & tf   \\
NGC4632           & 124232.8$-$000447 &1569 &   6 & 9.4 & NGC4666  &31.06   &       16.2   & tf   \\
MCG-02-32-26      & 124248.9$-$122326 & 828 &$-$1 &11.4 & NGC4594  &  $-$   &        $-$   &      \\
SHOC 381          & 124318.5$-$012803 &1529 &   8 &14.9 & NGC4666  &  $-$   &        $-$   &      \\
DDO 142           & 124403.5$-$054034 &1259 &   8 &11.7 & NGC4697  &  $-$   &        $-$   &      \\
UGC07911          & 124428.8$+$002805 &1041 &   9 &12.3 & NGC4753  &31.03   &       16.1   & tf   \\
UGC07913          & 124433.2$-$021909 &1431 &  10 &13.6 & NGC4666  &31.51   &       20.0   & tf   \\
VLA ...           & 124445.6$-$002536 &1225 &  10 &16.2 & NGC4666  &  $-$   &        $-$   &      \\
UGCA 295          & 124454.1$-$090731 &1194 &   8 &11.6 & NGC4697  &  $-$   &        $-$   &      \\
NGC4666           & 124508.7$-$002743 &1367 &   5 & 7.0 & NGC4666  &31.28   &       18.2   & tf   \\
NGC4668           & 124532.0$-$003209 &1476 &   7 &10.6 & NGC4666  &31.07   &       16.4   & tf   \\
DDO 146           & 124541.4$-$060408 &1303 &   9 &11.8 & NGC4697  &30.92   &       15.3   & tf   \\
SDSS ...          & 124541.6$-$020151 &1423 &  10 &13.6 & NGC4666  &  $-$   &        $-$   &      \\
SDSS ...          & 124547.9$-$002556 &1507 &   9 &14.0 & NGC4666  &  $-$   &        $-$   &      \\
NGC4674           & 124603.5$-$083920 &1325 &   1 &10.3 & NGC4697  &  $-$   &        $-$   &      \\
KDG 198           & 124637.5$-$040433 &1111 &  10 &14.3 & NGC4697  &30.71   &       13.9   & tf   \\
PGC043108         & 124649.3$+$030023 &1069 &  10 &13.0 & NGC4753  &  $-$   &        $-$   &      \\
NGC4684           & 124717.5$-$024340 &1420 &$-$1 & 8.4 & NGC4666  &30.65   &       13.5   & sbf  \\
PGC1004122        & 124724.9$-$082431 &1041 &   9 &13.8 & NGC4697  &31.89   &       23.8   & tf   \\
MCG-01-33-007     & 124738.2$-$055203 &1170 &   1 &11.2 & NGC4697  &  $-$   &        $-$   &      \\
PGC1003283        & 124750.9$-$082816 & 860 &$-$1 &12.1 & NGC4594  &  $-$   &        $-$   &      \\
NGC4691           & 124813.6$-$031958 & 954 &   0 & 8.5 & NGC4753  &  $-$   &        $-$   &      \\
HIPASSJ1248-08    & 124830.6$-$080232 &1325 &   9 &12.5 & NGC4697  &30.71   &       13.9   & tf   \\
NGC4697           & 124835.9$-$054803 &1071 &$-$5 & 6.4 & NGC4697  &30.49   &       12.5   & sbf  \\
DDO 148           & 124843.1$-$051514 &1175 &  10 &12.5 & NGC4697  &29.77   & \enskip9.0   & tf   \\
LCRS B...         & 124854.2$-$114035 &1145 &   9 &12.8 & NGC4697  &31.46   &       19.6   & tf   \\
NGC4699           & 124902.2$-$083953 &1218 &   3 & 6.5 & NGC4697  &31.55   &       20.4   & tf   \\
NGC4700           & 124908.2$-$112435 &1219 &   5 & 9.8 & NGC4697  &29.06   & \enskip7.3   & tf   \\
DDO 149           & 124918.3$-$040059 &1353 &   8 &12.4 &          &31.70   &       21.9   & tf   \\
MCG-02-33-015     & 124923.7$-$100706 &1131 &   7 &11.4 & NGC4697  &31.10   &       16.6   & tf   \\
NGC4678           & 124941.9$-$043447 &1269 &   8 &12.5 & NGC4697  &30.46   &       12.3   & tf   \\
UGC07982          & 124950.2$+$025110 &1024 &   4 &10.2 & NGC4753  &31.79   &       22.8   & tf   \\
PGC1019240        & 124955.9$-$072527 &1236 &   7 &12.2 & NGC4697  &  $-$   &        $-$   &      \\
MGC 0038179       & 125004.7$-$001357 & 608 &$-$1 &17.1 &          &  $-$   &        $-$   &      \\
KDG206            & 125007.3$+$021452 & 934 &$-$5 &13.2 & NGC4753  &  $-$   &        $-$   &      \\
IC 0825           & 125019.2$-$052147 &1367 &$-$0 &12.9 & NGC4697  &  $-$   &        $-$   &      \\
NGC4720           & 125042.8$-$040921 &1452 &   2 &10.8 &          &31.53   &       20.2   & tf   \\
NGC4731           & 125101.1$-$062335 &1325 &   6 & 9.8 & NGC4697  &30.66   &       13.6   & tf   \\
NGC4723           & 125102.9$-$131412 &1145 &   9 &12.4 & NGC4722  &30.93   &       15.3   & tf   \\
NGC4731A          & 125113.3$-$063329 &1327 &  10 &11.5 & NGC4697  &  $-$   &        $-$   &      \\
NGC4722           & 125132.4$-$131948 &1118 &   3 & 9.4 & NGC4722  &30.92   &       15.3   & tf   \\
NGC4742           & 125148.0$-$102717 &1086 &$-$5 & 8.4 & NGC4697  &30.95   &       15.5   & sbf  \\
SDSS ...          & 125208.9$+$030715 & 900 &$-$1 &15.2 & NGC4753  &  $-$   &        $-$   &      \\
NGC4753           & 125222.1$-$011159 &1091 &$-$1 & 6.7 & NGC4753  &31.08   &       16.9   & SN   \\
SDSS ...          & 125233.8$-$014349 & 983 &   9 &12.9 & NGC4753  &  $-$   &        $-$   &      \\
DDO 152           & 125236.3$-$061720 &1363 &   8 &12.8 & NGC4697  &31.23   &       17.6   & tf   \\
NGC4757           & 125250.1$-$101837 & 664 &   3 &10.3 &          &  $-$   &        $-$   &      \\
NGC4771           & 125321.2$+$011609 & 992 &   6 & 9.0 & NGC4753  &31.45   &       19.5   & tf   \\
UGCA 305          & 125321.4$-$045841 &1248 &  10 &11.9 & NGC4697  &31.21   &       17.5   & tf   \\
NGC4772           & 125329.2$+$021006 & 907 &   1 & 8.3 & NGC4753  &31.97   &       24.8   & tf   \\
$[$KKS2000$]$38   & 125331.6$-$055540 & 905 &  10 &14.0 &          &30.09   &       10.4   & tf   \\
$[$KKS2000$]$39   & 125341.8$-$060502 &1303 &  10 &14.2 & NGC4697  &  $-$   &        $-$   &      \\
NGC4775           & 125345.7$-$063720 &1398 &   7 & 9.2 & NGC4697  &  $-$   &        $-$   &      \\
UGCA 307          & 125357.5$-$120631 & 636 &   9 &11.8 &          &29.50   &\enskip8.0   & tf   \\
CGCG 015-033      & 125401.9$-$013030 & 985 &$-$1 &12.0 & NGC4753  &  $-$   &        $-$   &      \\
SDSS ...          & 125405.2$-$000604 & 697 &   9 &13.3 & NGC4753  &  $-$   &        $-$   &      \\
CGCG 015-035      & 125412.6$+$004809 &1033 &   9 &13.7 & NGC4753  &  $-$   &        $-$   &      \\
NGC4781           & 125423.7$-$103214 &1078 &   7 & 8.6 & NGC4697  &30.95   &       15.5   & tf   \\
SDSS ...          & 125437.7$+$011933 &1061 &   9 &15.0 & NGC4753  &  $-$   &        $-$   &      \\
NGC4790           & 125451.9$-$101452 &1173 &   5 & 9.8 & NGC4697  &31.14   &       16.9   & tf   \\
PGC0158179        & 125503.7$-$062210 &1141 &   8 &13.2 & NGC4697  &30.59   &       13.1   & tf   \\
2dFGRS...         & 125510.8$-$032351 &1329 &  10 &14.4 &          &30.95   &       15.5   & tf   \\
UGC08041          & 125512.7$+$000700 &1212 &   7 &12.0 & NGC4753  &31.17   &       17.2   & tf   \\
UGCA 308          & 125531.0$-$102353 &1141 &  10 &13.3 & NGC4697  &31.10   &       16.6   & tf   \\
2MASX ...         & 125537.4$-$011948 & 760 &   9 &13.6 &          &  $-$   &        $-$   &      \\
UGC08048          & 125539.4$-$001549 & 972 &   8 &10.9 & NGC4753  &31.35   &       18.7   & tf   \\
NGC4802           & 125549.6$-$120319 & 827 &$-$2 & 8.5 & NGC4594  &30.31   &       11.5   & sbf  \\
NGC4813           & 125636.1$-$064904 &1206 &$-$2 & 9.9 & NGC4697  &  $-$   &        $-$   &      \\
IC3908            & 125640.4$-$073340 &1133 &   5 & 9.1 & NGC4697  &31.90   &       23.9   & tf   \\
NGC4818           & 125648.9$-$083131 & 927 &   2 & 7.9 & NGC4594  &30.43   &       12.2   & tf   \\
UGCA 310          & 125712.1$-$040932 &1377 &   8 &12.7 &          &32.41   &       30.4   & tf   \\
MCG-01-33-059     & 125716.5$-$052045 &1097 &   8 &11.9 & NGC4697  &31.64   &       21.3   & tf   \\
UGCA 311          & 125746.8$-$093801 &1310 &   7 &10.1 & NGC4697  &31.49   &       19.8   & tf   \\
NGC4845           & 125801.2$+$013433 & 979 &   2 & 7.8 & NGC4753  &31.81   &       23.1   & tf   \\
MCG-02-33-075     & 125828.3$-$103437 &1095 &   8 &11.8 & NGC4697  &29.70   & \enskip8.7   & tf   \\
MCG-01-33-061     & 125848.9$-$060646 &1442 &   8 &12.2 &          &31.59   &       20.8   & tf   \\
APMUKS(BJ)...     & 125849.6$-$045319 &1155 &   9 &13.2 & NGC4697  &29.07   & \enskip6.5   & tf   \\
UGCA 312          & 125906.8$-$121340 &1122 &   8 &12.4 &          &30.39   &       12.0   & tf   \\
NGC4856           & 125921.3$-$150232 &1183 &   1 & 7.4 & NGC4856  &31.89   &       24.0   & tf   \\
SDSS ...          & 125945.1$-$005217 &1139 &   9 &15.2 & NGC4753  &  $-$   &        $-$   &      \\
SDSS ...          & 125952.4$+$033759 &1006 &   9 &15.7 & NGC4753  &  $-$   &        $-$   &      \\
KK176             & 125956.3$-$192447 & 620 &  10 &13.7 &          &29.19   & \enskip6.9   & tf   \\
MCG-02-33-082     & 130005.1$-$152155 &1260 &   9 &11.4 & NGC4856  &31.19   &       17.3   & tf   \\
UGCA 314          & 130017.0$-$122048 &1397 &   8 &11.3 &          &31.97   &       24.8   & tf   \\
SDSS ...          & 130017.6$-$030359 & 921 &   9 &13.4 & NGC4753  &  $-$   &        $-$   &      \\
DDO 159           & 130043.5$-$154256 &1189 &  10 &13.4 & NGC4856  &  $-$   &        $-$   &      \\
NGC4904           & 130058.7$-$000140 &1036 &   6 & 9.5 & NGC4753  &31.50   &       20.0   & tf   \\
UGC08127 n1       & 130100.7$-$015834 &1216 &  10 &13.6 &          &  $-$   &        $-$   &      \\
UGC08127          & 130103.7$-$015712 &1331 &   8 &12.7 &          &31.94   &       24.4   & tf   \\
APMUKS(BJ)...     & 130105.2$-$052821 & 936 &   9 &13.8 & NGC4941  &  $-$   &        $-$   &      \\
HIPASS1300-13B    & 130107.0$-$133106 &1121 &   9 &13.5 & NGC4856  &30.81   &       14.6   & tf   \\
MRK 1342          & 130110.9$-$053324 &1497 &   1 &11.3 &          &  $-$   &        $-$   &      \\
MCG-02-33-093     & 130203.3$-$145258 &1203 &   5 &11.0 & NGC4856  &31.88   &       23.7   & tf   \\
NGC4920           & 130204.1$-$112242 &1145 &   9 &11.6 &          &31.30   &       18.2   & tf   \\
UGCA 319          & 130214.4$-$171415 & 548 &  10 &11.6 & UGCA 320 &29.32   & \enskip7.3   & mem  \\
NGC4928           & 130300.6$-$080506 &1553 &   4 &10.2 & NGC4995  &31.32   &       18.4   & tf   \\
UGCA 320          & 130316.7$-$172523 & 546 &  10 &10.8 & UGCA 320 &29.32   & \enskip7.3   & tf   \\
2MASX ...         & 130412.1$-$045328 & 765 &$-$1 &11.0 &          &  $-$   &        $-$   &      \\
NGC4941           & 130413.1$-$053306 & 951 &   2 & 8.2 & NGC4941  &31.63   &       21.2   & tf   \\
NGC4942           & 130419.1$-$073858 &1586 &   7 &10.4 & NGC4995  &31.30   &       18.2   & tf   \\
UGCA 322          & 130431.2$-$033421 &1219 &   8 &11.7 &          &30.65   &       13.5   & tf   \\
LCRS B...         & 130431.8$-$025917 &1148 &  10 &13.6 & LCRS B13 &  $-$   &        $-$   &      \\
HIPASSJ1304-02    & 130446.6$-$025216 &1122 &  10 &14.3 & LCRS B13 &  $-$   &        $-$   &      \\
NGC4948           & 130456.0$-$075652 &1113 &   7 &10.7 & NGC4697  &30.88   &       15.0   & tf   \\
PGC0986100        & 130456.1$-$094849 &1288 &   9 &13.7 &          &  $-$   &        $-$   &      \\
NGC4948A          & 130505.8$-$080941 &1387 &   8 &11.9 &          &30.97   &       15.6   & tf   \\
NGC4951           & 130507.7$-$062938 &1016 &   6 & 8.9 & NGC4941  &30.82   &       15.1   & tf   \\
DDO 163           & 130514.3$-$075321 & 958 &   7 &10.5 & NGC4697  &31.87   &       23.7   & tf   \\
NGC4958           & 130548.9$-$080113 &1058 &$-$2 & 7.6 & NGC4697  &  $-$   &        $-$   &      \\
DDO 164           & 130618.4$-$173053 &1265 &  10 &13.0 &          &31.30   &       18.2   & tf   \\
PGC0158522        & 130701.7$-$074155 &1449 &$-$1 &12.3 &          &  $-$   &        $-$   &      \\
MCG-03-34-002     & 130756.7$-$164121 & 728 &   3 &12.5 &          &29.49   & \enskip7.9   & tf   \\
NGC4981           & 130848.7$-$064639 &1519 &   4 & 8.5 & NGC4995  &31.90   &       24.1   & tf   \\
NGC4984           & 130857.2$-$153059 &1070 &$-$1 & 7.7 &          &  $-$   &        $-$   &      \\
NGC4995           & 130940.7$-$075000 &1605 &   3 & 8.2 & NGC4995  &32.03   &       25.5   & tf   \\
UGCA 330          & 130947.5$-$101912 &1042 &   7 &10.0 &          &31.73   &       22.2   & tf   \\
PGC0984591        & 131038.6$-$095554 & 996 &  10 &14.0 &          &  $-$   &        $-$   &      \\
UGCA 332          & 131158.9$-$120349 &1932 &   7 &12.0 &          &32.38   &       30.0   & tf   \\
IC4212            & 131203.0$-$065929 &1328 &   6 &10.2 &          &31.70   &       21.9   & tf   \\
MCG-03-34-019     & 131305.5$-$162842 &1778 &   3 &10.5 & NGC5054  &  $-$   &        $-$   &      \\
LEDA 083827       & 131431.2$-$162248 &1311 &  10 &12.7 &          &  $-$   &        $-$   &      \\
NGC5035           & 131449.2$-$162934 &1993 &$-$1 & 9.6 & NGC5044  &  $-$   &        $-$   &      \\
PGC0083842        & 131449.8$-$165825 &1994 &$-$1 &11.7 & NGC5044  &  $-$   &        $-$   &      \\
NGC5037           & 131459.4$-$163525 &1698 &   3 & 8.6 & NGC5054  &  $-$   &        $-$   &      \\
2MASX ...         & 131504.1$-$162340 &1643 &   1 &12.2 & NGC5054  &  $-$   &        $-$   &      \\
MCG-03-34-40      & 131656.2$-$163535 &1953 &   7 &12.6 & NGC5044  &  $-$   &        $-$   &      \\
NGC5054           & 131658.5$-$163805 &1556 &   4 & 7.6 & NGC5054  &31.18   &       17.2   & tf   \\
UM 559            & 131742.8$-$010001 &1106 &  10 &13.7 &          &  $-$   &        $-$   &      \\
DDO 171           & 131841.2$-$082647 &1150 &  10 &11.6 &          &  $-$   &        $-$   &      \\
NGC5088           & 132020.2$-$123418 &1261 &   4 &10.1 &          &31.71   &       22.0   & tf   \\
ESO576-037        & 132028.9$-$195032 &1558 &   9 &13.8 & NGC5084  &  $-$   &        $-$   &      \\
MCG-03-34-054     & 132041.0$-$165402 &1518 &   5 &10.9 & NGC5054  &  $-$   &        $-$   &      \\
NGC5099           & 132119.6$-$130232 &1145 &$-$5 &12.0 &          &  $-$   &        $-$   &      \\
MCG-03-34-067     & 132416.2$-$164213 &1307 &   7 &12.3 & NGC5170  &  $-$   &        $-$   &      \\
UGCA 353          & 132442.1$-$194150 &1784 &   5 &10.6 &          &31.28   &       18.0   & tf   \\
HIPASSJ1325-12    & 132509.3$-$122544 &1003 &   7 &12.5 &          &  $-$   &        $-$   &      \\
HIPASSJ1326-09    & 132550.3$-$093209 &1651 &   9 &14.1 &          &  $-$   &        $-$   &      \\
PGC0990590        & 132636.6$-$092828 &1063 &$-$1 &13.0 &          &  $-$   &        $-$   &      \\
6dF ...           & 132726.3$-$173918 &1464 &   8 &12.8 & NGC5170  &  $-$   &        $-$   &      \\
MCG-03-34-082     & 132847.0$-$173044 &1267 &   4 &10.2 & NGC5170  &  $-$   &        $-$   &      \\
NGC5170           & 132948.8$-$175759 &1319 &   5 & 7.6 & NGC5170  &31.84   &       23.4   & tf   \\
KDG227            & 133439.7$-$121950 &1347 &   8 &12.2 &          &30.47   &       12.4   & tf   \\
\end{longtable}
\end{center}

\begin{center}
\tabcolsep=0.6ex%
\begin{longtable}{lcrrrrrlrrcllr}
\caption{MK-groups in the region of Virgo Southern Extension}\\
\hhline{==============}
Name  &RA(J2000.0)Dec.& $N_v$ & $V_{LG}$&$\sigma_v$ & $R_h$  &$lg L_K$  &$\lg M_{\odot}$   &$\frac{M_{\odot}}{L_{\odot}}$ & $T_1$&$<m$-$M>$  & $\sigma_{m-M}$  & ~$D$   & $n_D$ \\
\hhline{==============}\endfirsthead
\hhline{==============}
Name  &RA(J2000.0)Dec.& $N_v$ & $V_{LG}$&$\sigma_v$ & $R_h$  &$lg L_K$  &$\lg M_{\odot}$   &$\frac{M_{\odot}}{L_{\odot}}$ & $T_1$&$<m$-$M>$  & $\sigma_{m-M}$  & ~$D$   & $n_D$ \\
\hhline{==============}\endhead
\hhline{--------------}\endfoot
\hhline{==============}\endlastfoot
N4437  &123310.7$+$000446&  3 & 947 & 23 &290& 10.40            &       11.61&  16&    9&  30.01& ~0.12& 10.0&  3  \\
N4546  &123529.5$-$034735&  4 & 884 & 34 & 91& 10.52            &       11.73&  16& $-$3&  30.38& ~0.37& 11.9&  2  \\
N4594  &124112.5$-$112103& 16 & 879 & 80 &642& 11.59            &       13.17&  38&    1&  30.52& ~0.52& 12.7& 11  \\
N4666  &124518.9$-$005318& 14 &1453 & 98 &251& 11.20            &       12.86&  46&    5&  31.11& ~0.28& 16.7&  5  \\
N4722  &125130.7$-$131929&  2 &1131 & 14 & 31&  9.96            &       10.50&   4&    3&  30.92& ~0.4 & 15.3&  2  \\
N4697  &125148.6$-$074321& 37 &1196 &107 &506& 11.59            &       13.23&  44& $-$5&  30.87& ~0.75& 14.9& 26  \\
N4753  &125326.6$-$001502& 23 & 984 & 97 &677& 11.54            &       12.96&  26& $-$2&  31.47& ~0.30& 19.7& 10  \\
N4856  &125929.5$-$150142&  5 &1191 & 44 &258& 10.94            &       12.22&  30&    0&  31.44& ~0.46& 19.4&  4  \\
UA320  &130256.5$-$172146&  2 & 547 &  1 & 40&  8.9\enskip\strut& \enskip8.5&    1&   10&  29.32& ~0.4 &  7.3&  1  \\
N4941  &130413.1$-$053306&  3 & 968 & 35 &270& 10.71            &       12.08&  23&    2&  31.22& ~0.40& 17.5&  2  \\
N4995  &130831.2$-$072620&  4 &1566 & 33 &420& 11.00            &       11.91&   8&    3&  31.64& ~0.33& 21.3&  4  \\
N5054  &131632.6$-$163759&  5 &1639 & 94 & 60& 10.97            &       12.80&  65&    4&  31.18& ~0.4 & 17.2&  1  \\
N5170  &132938.8$-$175442&  4 &1339 & 75 &290& 11.05            &       12.62&  35&    5&  31.84& ~0.4 & 23.3&  1  \\
\end{longtable}
\end{center}


\begin{thebibliography}{}
\bibitem{}Abazajian K.N., Adelman-McCarthy J.K., Agueros M.A., et al. 2009, ApJS, 182, 54
\bibitem{}Binggeli B., Popescu C.C., Tammann G.A., 1993, A \& A Suppl, 98, 275
\bibitem{}Chernin A.D., Teerikorpi P., Valtonen M. J., et al. 2012 A \& A, 539, 4
\bibitem{}de Vaucouleurs G., de Vaucouleurs A., 1973, A \& A, 28, 109
\bibitem{}Dietrich J.P., Werner N., Clowe D. et al. 2012, Nature (arXiv:1207.0809)
\bibitem{}Dinshaw, N., Weymann, R. J., Impey, C. D.,  et al. 1997, ApJ, 491, 45
\bibitem{}Fukugita M., Peebles P.J.E., 2004, ApJ, 616, 643
\bibitem{}Heisler J., Tremaine S., Bahcall J.N., 1985, ApJ, 298, 8
\bibitem{}Impey C.D., Petry C.E., Flint K.P., 1999, ApJ, 524, 536
\bibitem{}Jarrett T.N., Chester T., Cutri R. et al. 2000, AJ, 119, 2498
\bibitem{}Jones D.H., Peterson B.A., Colless M., Daunders W., 2006, MNRAS, 369, 25
\bibitem{}Karachentsev I.D., Makarov D.I., Kaisina E.I., 2013, AJ (submitted)
\bibitem{}Karachentsev I.D., Nasonova O.G., Courtois H.M., 2012, MNRAS (accepted; arXiv:1211.5975)
\bibitem{}Karachentsev I.D., 2012, Astrophys. Bull., 67, 123
\bibitem{}Karachentsev I.D., Nasonova O.G., Courtois H.M., 2011, ApJ, 743, 123
\bibitem{}Karachentsev I.D., Nasonova O.G., 2010, MNRAS, 405, 1075
\bibitem{}Karachentsev I.D., Makarov D.I., 2008, Astrophys. Bulletin, 63, 299
\bibitem{}Lauberts A., Valentijn E.A., 1985, Photographic photometry of 16000 galaxies
on ESO blue and red survey plates, Springer-Verlag
\bibitem{}McLaughlin D.E., 1999, ApJ, 512, L9
\bibitem{}Magtesian A., 1988, Astrofizika, 28, 150
\bibitem{}Makarov D.I., Karachentsev I.D., 2011, MNRAS, 412, 2498
\bibitem{}Makarov D.I., Karachentsev I.D., 2009, Astrophys. Bulletin, 64, 24
\bibitem{}Masaki S., Fukugita M., Yoshida N., 2011, (arXiv:1105.3005)
\bibitem{}Mei S., et al., 2007, ApJ, 655, 144
\bibitem{}Meyer M.J., Zwaan M.A., Webster R.L. et al., 2004, MNRAS, 350, 1195
\bibitem{}Rosenberg J.L., Ganguly R., Giroux M.L., Stoke J.T., 2003, ApJ, 591, 677
\bibitem{}Spergel D.N. et al. 2007, ApJS, 170, 377
\bibitem{}Springob C.M., Masters K.L., Haynes M.P. et al. 2009, ApJS, 172, 599
\bibitem{}Tavio H., Cuesta A.J., Prada F. et al., 2008 (arXiv:0807.3027)
\bibitem{}Tonry J.L., Dressler A., Blakeslee J.P., et al. 2001, ApJ, 546, 681
\bibitem{}Tully R.B., Rizzi L., Shaya E.J., et al. 2009, AJ, 138, 323
\bibitem{}Tully R.B., 1987, ApJ, 321, 280
\bibitem{}Tully R.B., Shaya E.J., 1984, ApJ, 281, 31
\bibitem{}Tully R.B., 1982, ApJ, 257, 389
\bibitem{}Tully R.B., Fisher R.J., 1977, A \& A, 54, 661
\bibitem{}Vennik J., 1984, Tartu Astron. Obs. Publ., 73, 1
\bibitem{}Yoon J.H., Putman M.E., Thom C. et al. 2012 (arXiv:1204.631)
\bibitem{}Zwaan M.A., Staveley-Smith L., Koribalski B.S. et al., 2003, AJ, 125, 2842
\bibitem{}Zwicky F., Herzog E., Wild P., Karpowich M., 1961-1968, Catalogue of
 Galaxies and Cluster of Galaxies, Pasadena, California Institute of
Technology
\end{thebibliography}
\end{document}